\documentclass[preprint,aps,double-space]{revtex4}

\usepackage{graphicx}
\usepackage{dcolumn}
\usepackage{bm}

\begin{document}

\title{Damping of Condensate Oscillation of a Trapped Bose Gas in a One-Dimensional Optical Lattice at Finite Temperatures}

\author{Emiko Arahata}
\author{Tetsuro Nikuni}%
\affiliation{%
Department physics, Faculty of science, Tokyo University of Science, \\
1-3 Kagurazaka, Shinjuku-ku, Tokyo 162-8601, Japan}%
\date{\today}

\begin{abstract}
We study damping of a dipole oscillation in a Bose-Condensed gas in a combined cigar-shaped harmonic trap and one-dimensional (1D) optical lattice potential at finite temperatures. In order to include the effect of thermal excitations in the radial direction, we derive a quasi-1D model of the Gross-Pitaeavskii equation and the Bogoliubov equations. 
We use the Popov approximation to calculate the temperature dependence of the condensate fraction with varying lattice depth.
We then calculate the Landau damping rate of a dipole oscillation as a function of the lattice depth and temperature. 
The damping rate increases with increasing lattice depth, which is consistent with experimental observations. The magnitude of the damping rate is in reasonable agreement with experimental data. We also find that the damping rate has a strong temperature dependence, showing a sharp increase with increasing temperature. Finally, we emphasize the importance of the radial thermal excitations in both equilibrium properties and the Landau damping.    
\pacs{03.75.Kk, 03.75.Lm, 67.40.-w}\end{abstract}
\maketitle
\section{Introduction}
Recently the dynamics of ultracold atomic gases in optical lattices have attracted attention both theoretically 
and experimentally \cite{RMP78}. In particular, center-of-mass dipole oscillations of Bose-Einstein condensates in a combined cigar shaped trap and one dimensional (1D) optical lattice potential have been experimentally studied in detail \cite{PRL86,Sc293,PRA66}.
In the presence of the periodic lattice potential, a decrease of the dipole mode frequency was observed \cite{Sc293}. This decrease can be understood in terms of the increase of the effective mass due to the lattice potential \cite{EPJD27}. On the other hand, at finite temperatures where an appreciable number of atoms are thermally excited out of the condensate, strong damping of the dipole oscillation of the condensate  was observed in the presence of the lattice potential \cite{PRA66}. In a pure harmonic potential, a Bose gas exhibits undamped dipole oscillations even at finite temperatures, since the condensate and noncondensate atoms oscillate with the same frequency without changing their density profiles \cite{JLTP116}. In contrast, in the presence of the periodic lattice potential, only the condensate component can coherently tunnel through the potential barriers, while the thermal component is locked by the lattice potential \cite{PRA66}. It is clear that this incoherent thermal conponent gives rise to the damping of the condensate oscillations. In Bose-condensed gases trapped in harmonic potentials, Landau damping is known to be the dominant contribution to damping of the condensate oscillations in the collisionless regime \cite{PRA57,PRL235,PRA61,NJP5}. 
In the analysis of damping of dipole oscillations in optical lattice, the authors of Ref.~\cite{PRA66} gave a rough estimate of the Landau damping rate and compared it with their experimental data. 
However, quantitative calculations of the Landau damping rate of dipole condensate oscillations in optical lattices have not given in any detail so far. In fact, even equilibrium properties of a Bose gas in an optical lattice at finite temperatures in connection with the experimentas of Ref.~\cite{PRA66} have not  been studied in detail.  
Although several papers have discussed finite-temperature properties of ultracold atoms in a 1D optical lattice, most theoretical studies have concentrated on the first Bloch band using the Bose-Hubbard model, and have ignored the effect of radial excitations \cite{PRA 70,PRA 72,PRA 73}. This approximation is only valid when $k_{\rm{B}}T\ll E_{\rm{R}}$ and $k_{\rm{B}}T\ll \hbar\omega_\bot$, where $\it{E}_{\rm{R}}$ is the recoil energy that is roughly the highest energy in the first Bloch band, and $\hbar\omega_\bot$ is the first excitation energy in the radial direction. In order to obtain more quantitative results that is applicable to the experiments of Refs.~\cite{PRA66,Sc293}, however, it is important to consider thermal excitations in the higher bloch bands as well asa the radial direction, since the experiments do not always satisfy $k_{\rm{B}}T\ll E_{\rm{R}}$ and $k_{\rm{B}}T\ll \hbar\omega_\bot$.    \par 
In this paper, we study the Landau damping of condensate oscillations of a trapped Bose gas in a 1D optical lattice, with explicitly including the effect of the radial excitations. In Sec.~II, we derive a quasi-1D model of the Gross-Pitaeavskii equation and the Bogoliubov equations that include the effect of the excitations in the radial direction. Using the Hatree-Fock-Bogoliubov-Popov (HFB-Popov) approximation, we solve these equations to calculate the temperature dependence of the condensate fraction. 

In Sec.~III, we calculate the Landau damping of the dipole condensate oscillation in an optical lattice. We calculate the damping rate as a function of the lattice depth with a fixed temperature, and compare it with the experimental data \cite{PRA66}. 
The magnitude of the damping rate is found to be in reasonable agreement of the experimental data \cite{PRA66}. The damping rate increases with increasing lattice depth, which is consistent with the experimental result \cite{PRA66}. We also calculate the temperature dependence of damping rate with a fixed lattice depth. We find that the damping has a strong temperature, showing a sharp increase with increasing temperature.
\section{Quasi 1D modeling of a trapped Bose gas}
We consider a Bose condensed gas in a combined potential of highly-elongated harmonic trap and 1D optical lattice.
Our system is described by the following Hamiltonian:
\begin{eqnarray}
\hat{H}=\int d\textbf{r}\Bigg\{\hat{\psi}^\dagger (\textbf{r})\left[-\frac{\hbar^2}{2m}\nabla ^2+V_{\rm{ext}}(\textbf{r})\right]\hat{\psi }(\textbf{r})\nonumber \\ 
+\frac{g}{2}\hat{\psi}^\dagger(\textbf{r})\hat{\psi}^\dagger (\textbf{r})\hat{\psi }(\textbf{r})\hat{\psi }(\textbf{r})\Bigg\},\label{H}
\end{eqnarray}
where $g=\frac{4\pi\hbar^2a}{m}$ is the coupling constant determined by the $s$-wave scattering length $a$. The external potential $V_{\rm{ext}}$ is given by
$V_{\rm{ext}}(\textbf{r})=V_{\rm{trap}}(\textbf{r})+V_{\rm{op}}(z)$, where  
$V_{\rm{trap}}(\textbf{r})=\frac{m}{2}\left[\omega _\bot ^2(x^2+y^2)+\omega_z^2z^2\right]$ is an anisotropic harmonic potential and 
$V_{\rm{op}}(z)=sE_{\rm{R}}\cos^2(kz) $
 is an optical lattice potential.
 Here $s$ is a dimensionless parameter describing the intensity of the laser beam creating the 1D lattice in units of the recoil energy $E_{\rm{R}}\equiv\frac{\hbar^2k^2}{2m}$, where $k=\frac{2\pi}{\lambda}$ is fixed by the wavelength $\lambda$ of the laser beam.
\par In this paper we consider a highly-anisotropic cigar shaped harmonic trap potential $\omega _\bot \gg \omega _z $. In order to take into account this quasi-1D situation, we expand the field operator in terms of the radial wave function \cite{cd06}: 
\begin{equation}
\hat{\psi }(\textbf{r})=\sum _{\alpha }\hat{\psi }_{\alpha}(z)\phi _\alpha (x,y)\label{expand},
\end{equation}
where $\phi_\alpha(x,y) $ is the eigenfunction of the radial part of the single-particle Hamiltonian,
\begin{eqnarray}
\left[-\frac{\hbar^2}{2m}\nabla _{\bot}^2+\frac{m}{2}\omega _\bot ^2(x^2+y^2)\right]\phi _\alpha (x,y)\nonumber \\ =\epsilon _\alpha\phi _\alpha (x,y)\label{xy1},
\end{eqnarray}
which satisfy the orthonormality condition
$\int dxdy \ \phi ^\ast _\alpha (x,y)\phi _\beta  (x,y)=\delta _{\alpha \beta}\label{xy2}$.
Here $\alpha=(n_x,n_y)$ is the index of the single-particle state with the eigenvalue $\epsilon_{(n_x,n_y)}=\hbar\omega_\bot(n_x+n_y+1)$.
Inserting Eq.~(\ref{expand}) into Eq.~(\ref{H}) and using Eq.~(\ref{xy1}), we obtain
\begin{eqnarray}
\hat{H}=\sum _\alpha \int dz \hat{\psi }^\dag_{\alpha}(z)\Biggl[-\frac{\hbar^2}{2m}\frac{\partial ^2}{\partial z^2}+\frac{m}{2}\omega _z^2z^2+V_{\rm{op}}(z)+\epsilon _\alpha \Biggr]\hat{\psi }_{\alpha}(z)\nonumber\\
+\sum _{\alpha  \alpha ^\prime \beta \beta^\prime}\frac{g_{\alpha  \alpha ^\prime \beta \beta^\prime}}{2}\int dz \hat{\psi }_{\alpha}^\dagger \hat{\psi }_{\beta}^\dagger \hat{\psi }_{\beta ^\prime}\hat{\psi }_{\alpha ^\prime},
\end{eqnarray}
where the renormalized coupling constant is defined by
\begin{equation}
g_{\alpha  \alpha ^\prime \beta \beta^\prime}\equiv g\int dxdy \phi _{\alpha}^\ast \phi _{\beta }^\ast \phi _{\beta ^\prime}\phi _{\alpha ^\prime}.
\end{equation}
\par Following the procedure described in Ref.~\cite{PRB53}
, we separate out the condensate wavefunction from the field operator as 
$
\hat{\psi}_\alpha=\langle\hat{\psi}_\alpha\rangle+\tilde{\psi}_\alpha \equiv \Phi_\alpha+\tilde{\psi}_\alpha 
$,
where $ \Phi_\alpha=\langle\hat{\psi}_\alpha\rangle$ is the condensate wavefunction and $\tilde{\psi}_\alpha$ is the noncondensate field operator.
Within the HFB-Popov approximation, we obtain the generalized Gross-Pitaevskii (GP) equation,
\begin{eqnarray}
&&\mu \Phi _\alpha=\left[-\frac{\hbar^2}{2m}\frac{\partial ^2}{\partial z^2}+\frac{m}{2}\omega _z^2z^2+V_{\rm{op}}+\epsilon _\alpha \right]\Phi _\alpha\nonumber\\
&&+\sum _{\alpha ^\prime \beta \beta^\prime}g_{\alpha  \alpha ^\prime \beta \beta^\prime}\left(\Phi^\ast _{\beta}\Phi _{\beta^\prime}\Phi _{\alpha^\prime}+2\Phi _{\beta^\prime} \left\langle\tilde{\psi}_{\beta}^\dag \tilde{\psi}_{\alpha ^\prime}\right\rangle  \right).
\label{GPal}
\end{eqnarray}
The Popov approximation neglects the anomalous correlation  $\left\langle \tilde{\psi}\tilde{\psi}  \right\rangle$ \cite{PRB53}. From the numerical solutions of the GP equation (\ref{GPal}) using the trap frequencies relevant to the experiment \cite{PRA66}, we find that $|\Phi_\alpha|^2/|\Phi_0|^2 \ll 10^{-6}$ ($\alpha\neq 0$), where we have denoted the lowest radial state as $\alpha=0=(0,0)$, and thus the contribution from higher radial modes to the condensate wavefunction is negligible small. For this reason, we will henceforth approximate $\Phi_\alpha\approx\Phi\delta_{\alpha,0}$.\par
Taking the usual Bogoliubov transformations for the noncondensate,
\begin{eqnarray}
\tilde{\psi}_\alpha(z)=\sum_{j}\left(u_{j\alpha}\hat{\alpha}_j-v_{j\alpha}^*\hat{\alpha}_j^\dag  \right),
\end{eqnarray}
 we obtain the coupled Bogoliubov equations,
 \begin{eqnarray}
 &\hat{L}_\alpha u_{j \alpha}+\sum_{\alpha^\prime} \Biggl[ \left( 2g^\alpha_{\alpha^\prime} n_0 +g_{\alpha\alpha^\prime \beta \beta^\prime}\tilde{n}_{\beta\beta^\prime} \right) u_{j \alpha^\prime}-g^\alpha_{\alpha^\prime} n_0 v_{j \alpha^\prime} \Biggr]
 \nonumber \\ &=E_j u_{j \alpha},\label{U}\\
& \hat{L}_\alpha v_{j \alpha}+\sum_{\alpha^\prime} \Biggl[ \left( 2g^\alpha_{\alpha^\prime} n_0 +g_{\alpha\alpha^\prime \beta \beta^\prime}\tilde{n}_{\beta\beta^\prime}\right) v_{j \alpha^\prime}-g^\alpha_{\alpha^\prime} n_0 u_{j \alpha^\prime} \Biggr]
\nonumber \\ &=-E_j u_{j \alpha},\label{V}
 \end{eqnarray}
 where we have introduce the operator 
  \begin{eqnarray}
   \hat{L}_\alpha&\equiv&-\frac{\hbar^2}{2m}\frac{\partial ^2}{\partial z^2}+\frac{m}{2}\omega _z^2z^2+V_{\rm{op}}(z)+\epsilon _\alpha-\mu .
 \end{eqnarray}
 We have also introduce the simplified notations 
 $n_0(z)=|\Phi(z)|^2,
   g^\alpha_{\alpha^\prime}=g_{\alpha\alpha^\prime00},
 \tilde{n}_{\beta\beta^\prime}=\left\langle\tilde{\psi}^\dag_\beta\tilde{\psi}_{\beta^\prime}  \right\rangle $. As noted above, we only include the  lowest mode ($\alpha=0$) in the condensation wavefunction $\Phi$.  
 Sums over the repeated indices $\beta, \beta^\prime$ are implied in Eqs.~(\ref{U}), and (\ref{V}).
 These equations define the quasi-particle excitation energies $E_j$ and the quasi-particle amplitudes $u_{j\alpha}$ and $u_{j\alpha}$. The orthonormality of the quasi-particle amplitudes is specified by the relation 
 \begin{eqnarray}
\int dz \left(u_{j\alpha}^\ast u_{i\alpha}-u_{i\alpha}u_{j\alpha}^\ast \right)=\delta_{ij}
\end{eqnarray}
 Using the solutions of Eqs.~(\ref{U}) and (\ref{V}), one can obtain the noncondensate density from 
$\tilde{n}=\sum_{\alpha}\tilde{n}_{\alpha \alpha}$,
where 
\begin{eqnarray}
\tilde{n}_{\alpha \beta}=\sum_{j}\biggl[\left(u_{j\alpha}u_{j\beta}+v_{j\alpha}v_{j\beta}\right)N(E_j)+v_{j\alpha}v_{j\beta}\biggr]\end{eqnarray}
with $N(E_j)=1/\left[ \exp(\beta E_j )-1\right]$.

Solving the coupled equations (\ref{GPal}), (\ref{U}) and (\ref{V}), we self-consistently determine the excitations spectrum $E_{\rm{j}}$ and the condensate fraction at finite temperatures.
Our calculation procedure is summarized as follows. Eq.~(\ref{GPal}) is first solved self-consistently for $\mu$ and $\Phi$ neglecting the interaction terms. Once $\Phi$ is known, $E_j$, $u_{j\alpha}$ and $u_{j\beta}$ are obtained from Eqs.~(\ref{U}) and (\ref{V}) with $\tilde{n}_{\beta\beta^\prime}$ set to zero. This is inserted into Eq.~(\ref{GPal}) and the process is repeated until convergence is reached. At each step, we define the normalization of the condensate wavefunctions by$\int dz \ |\Phi(z)|^2= N-\tilde{N}$, where $\tilde{N}=\int dz \ \tilde{n}(z)$ is the total number of noncondensate atoms.\par
\begin{figure}[htbp]
\centerline{\includegraphics[height=2.0in]{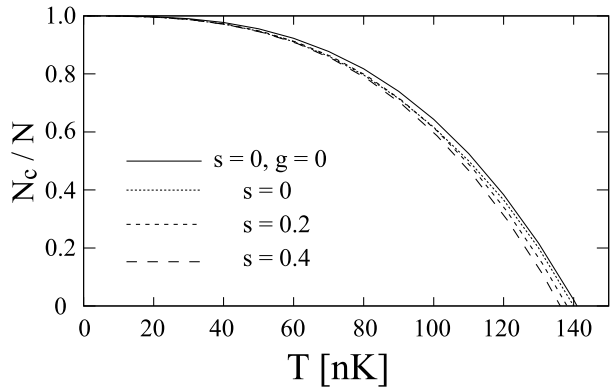}}
\caption {The condensate fraction $N_{\rm{c}}/$$N$ as a function of the temperature. $``s$=0" represents the ideal Bose gas result without a lattice potential $N_{\rm{c}}/N=1-(T/T^0_{\rm{c}})^3$, where $T^0_{\rm{c}}=$141 nK.}
\label{fig:Tc0}
\end{figure}

Throughout this paper we use the following parameters of the experiment of Ref.~\cite{PRA66}: $m(^{87}$Rb)=1.44$\times 10^{-25}$ kg, $\omega_z/2 \pi= 9.0$ Hz, $\omega_{\bot}/2\pi= 92$ Hz, scattering length $a=5.82$ nm and the wavelength of the optical lattice $\lambda$=795 nm. We fix the total number of atoms as $N=4\times 10 ^5$.
In Fig.~\ref{fig:Tc0}, we plot the condensate fraction $N_{\rm{c}}/$$N$ as a function of the temperature for various values of the lattice depth $s$ by solving GP equation Eq.~(\ref{GPal}) and Bogolibov equations Eqs.~(\ref{U}) and (\ref{V}).
It can be seen that each line falls to zero at approximately $T=$140 nK, which is close to  the semiclassical prediction of the BEC transition temperature of an ideal Bose gas in a 3D harmonic trap $T_{\rm{c}}^0$$= 0.94\hbar (\omega_\bot^2\omega_z)^{1/3}N^{1/3}/k_{\rm{B}}$= 141 nK \cite{p/s}.    In contrast,  $T_{\rm{c}}$ of an ideal Bose gas in a 1D harmonic trap is $T_{\rm{c}}$$ \sim \hbar\omega_z\frac{N}{k_{\rm{B}}\ln N}$=13.4 $\mu$K \cite{p/s}. This crearly shows that one must explicitly take into account the excitations in the radial direction in order to obtain the correct thermodynamic behavior at finite temperatures \cite{cd06}. The transition temperature deceases with increasing lattice depth, but there is no significant change in the temperature dependence of the condensate fraction, as long as the lattice potential is not so deep, i. e., $s\le 2$,
\par
\begin{figure}[htbp]
\centerline{\includegraphics[height=2.0in]{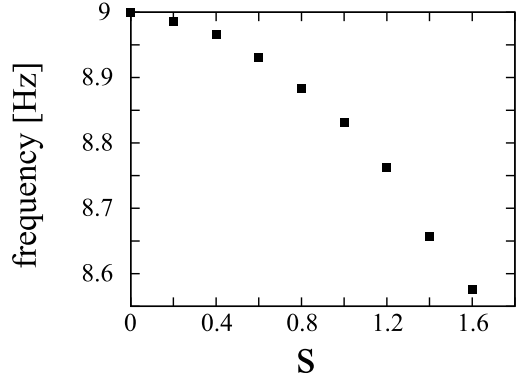}}
 \caption{The dipole mode frequency as a function of  the lattice depth $s$ with a fixed temperature $T=40$nK.}
 \label{fig:w}
\end{figure}

 We turn to the detailed structure of the excitation spectrum $E_j$.
From Eqs.~(\ref{U}) and (\ref{V}), one sees that the different radial modes are coupled due to interactions. However the coupling is so small that there are still distinct branches corresponding the radial modes. We thus label each branch with the index $\alpha$.For example, the lowest branch ($\alpha=0$) can be identified with the branch corresponding to the lowest radial mode.
\begin{figure}[htbp]
\centerline{\includegraphics[height=2.0in]{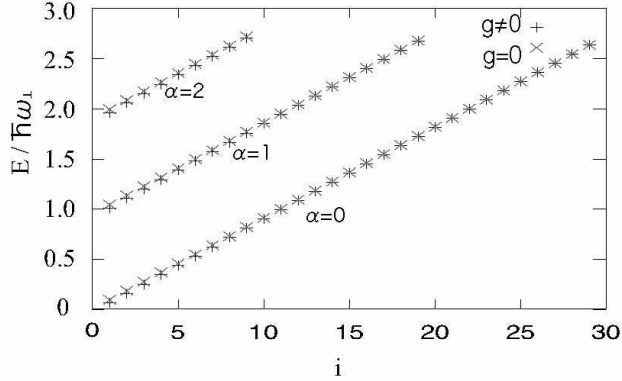}}
 \caption{The Bogoiubov excitation energy spectrum for $s$=1.2, where $i$ is the energy label for each branch } 
  \label{fig:ene}
\end{figure} 
In Fig~\ref{fig:ene}, we plot the excitation spectrum. Frequency of the condensate collective mode is related the excitation energy through $E_j=\hbar\omega_j$. In particular, the dipole mode frequency can be identified with the lowest frequency.  
In Fig.~\ref{fig:w}, we plot the dipole-mode frequency as a function of the lattice depth with a fixed temperature $T$= 40nK. One can see the decrease of the dipole-mode frequency with increasing lattice depth. The same behavior was also observed in the experiment \cite{Sc293}. The negative energy shift also found for other excitation modes. The decrease of $T_{\rm{c}}$ can be attributed to these negative shifts.       
\section{Landau damping rate of dipole mode}
In this Section, we calculate Landau damping of the dipole mode. Landau damping is the dominant damping mechanism for low-lying collective modes in trapped Bose-condensed gases in the collisionless regime at finite temperatures \cite{PRL235}.
Landau damping originates from the coupling between single-particle excitations and the collective oscillations. Damping occurs because the thermal bath of the elementary excitations can absorb quanta of the collective oscillations. 
A general expression for the Landau damping rate is given in Refs.~\cite{PRA57, PRL235,NJP5,p/s}. In our quasi-1D model, Landau damping rate can be expressed in terms of the quasi-particle excitation energies $E_j$ and quasi-particle amplitudes $u_{j\alpha}$ and $v_{j\alpha}$ calculated in the previous Section:
\begin{eqnarray}
\gamma_{\rm{L}}=4\pi\sum_{\alpha,\alpha^\prime}g_{\alpha\alpha^\prime 00}
\sum_{i\neq j}|A_{i j}^{\alpha\alpha^\prime}|^2\Bigl[N(E_i)-N(E_j)\Bigr]
\delta\left(\hbar\omega+E_i-E_j\right),\label{rate}
\end{eqnarray}
where $\omega$ is the eigenfrequency of a condensate dipole oscillation. The matrix element $A_{i j}^{\alpha\alpha^\prime}$ is defined by,
\begin{eqnarray}
A_{i j}^{\alpha\alpha^\prime}\equiv \int dz \ \Phi \{u_{10}\left[u_{i\alpha}^\ast 
u_{j\alpha^\prime}+v_{i\alpha}^\ast v_{j\alpha^\prime}-v_{i\alpha}^\ast u_{j\alpha^\prime}\right]\nonumber\\
-
v_{10}\left[u_{i\alpha}^\ast u_{j\alpha^\prime}+v_{i\alpha}^\ast v_{j\alpha^\prime}-u_{i\alpha}^\ast v_{j\alpha^\prime}\right]\}\label{rate_A}
\end{eqnarray}  
where $u_{10}$ and $v_{10}$ are the quasi-particle amplitudes corresponding the dipole mode. 
One difficulty in calculating the damping rate in Eq.~(\ref{rate}) using the discrete excitation spectrum $E_j$ is that it involves the energy-conserving delta functions. This difficulty can be overcome by replacing each delta function by a function with a finite width. In this paper, we use the following replacement: 
\begin{eqnarray}
\delta\left(\hbar\omega+E_i-E_j\right)\rightarrow\frac{1}{2\Delta}\Theta \left(\Big|\Delta-\hbar\omega+E_i-E_j\Big| \right).
\end{eqnarray}
\begin{figure}[htbp]
\centerline{\includegraphics[height=2.0in]{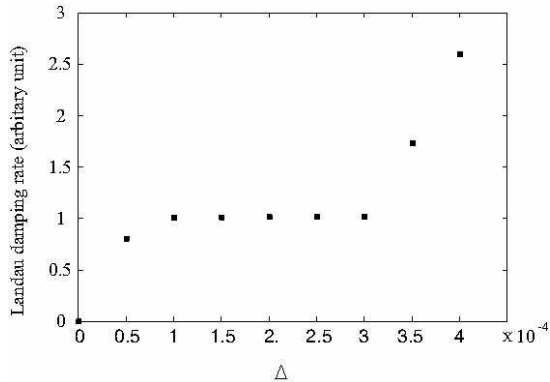}}
 \caption{The variation of the Landau damping rate $\gamma_{\rm{L}}$ (arbitrary unit) with the width factor $\Delta$ for lattice depth $s=1.0$ and temperature $T=120$ nK. To capture the $\Delta$ dependence of $\gamma_{\rm{L}}$ while saving computational time, we take the matrix element $A_{i j}^{\alpha\alpha^\prime}$ as a constant in this figure.} 
  \label{fig:delta}
\end{figure}
The width factor $\Delta$ is somewhat arbitrary, and the result for $\gamma_{\rm{L}}$ will vary with $\Delta$. In Fig.~\ref{fig:delta}, we plot a typical behavior of the $\Delta$ dependence of $\gamma_{\rm{L}}$. One can see  that the variation of $\gamma_{\rm{L}}$ is very weak when $\Delta/\hbar \omega_z$ lies between 1$\times10^{-4}$ and 3 $\times10^{-4}$. The same behaviors are also found for other temperatures and lattice heights. We thus take $\Delta/\hbar \omega_z=2.5\times10^{-4}$ in calculating the damping rate $\gamma_{\rm{L}}$.
\begin{figure}[htbp]
\centerline{\includegraphics[height=2.0in]{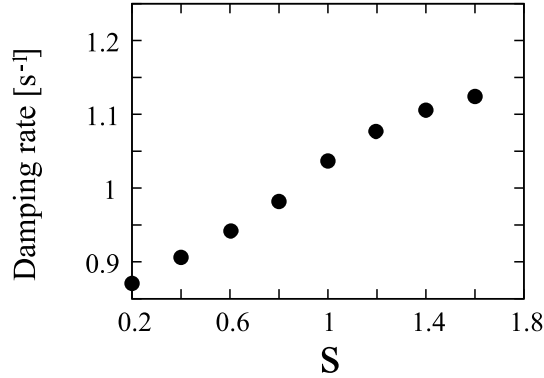}}
 \caption{The damping rate as a function of lattice depth $s$ with a fixed temperature $T=120$nK.} 
  \label{fig:rate}
\end{figure}
\begin{figure}[htbp]
\centerline{\includegraphics[height=2.0in]{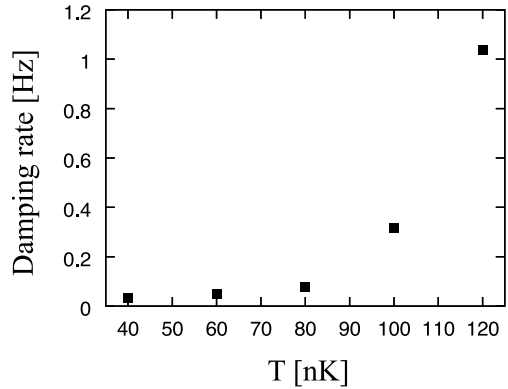}}
 \caption{The damping rate as a function of temperature with a fixed lattice depth $s=1.0$} 
  \label{fig:rate_co}
\end{figure}

In Fig.~\ref{fig:rate}, we plot the damping rate as a function of lattice depth $s$ with a fixed temperature $T=120$nK calculated from Eq.~(\ref{rate}). We see that the magnitude of the damping rate is $\sim1s^{-1}$, which is in reasonable agreement with the experimental data \cite{PRA66}. We also see that the damping rate increases with increasing lattice depth, which is consistent with the experimental result \cite{PRA66}. This increase of the damping rate can be attributed to the increase of the number of the elementary excitations because of a reduction of the excitation energy with increasing lattice depth. 
\par We next investigate the temperature dependence of the damping rate. In Fig.~{\ref{fig:rate_co}}, we plot the damping rate as a function of temperature with a fixed lattice depth $s=1.0$. We find that damping rate decreases rapidly with decreasing temperature. This is due to the reduction of thermal excitations with decreasing temperature. Because of this strong temperature dependence of the damping rate, it is very difficult to make a quantitative comparison with the experimental data without precise knowledge of the experimental temperature.
\par
Here we comment on the effect of adiabatic loading of an optical lattice on the temperature of a gas.
The usual path for preparing for condensed Bose gases in an optical lattice consists of first forming a ultracold bosons in a weak magnetic trap, to which a 1D lattice potential is adiabatically applied by slowly rating up the light field intensity. In this case, the initial and final temperature are not usually equal since the energy spectrum changes during lattice loading \cite{PRA69}. We used the entropy-temperature curves to consider the effect of the adiabatic loading into a lattice. 
For the initial ($s$=0) temperature $T$=120nK, the temperature shift in the final state $s=1.6$ is only about 2nK. Thus, one can ignore the effect of the adiabatic loading in shallow lattice regime.
 \par
We note that the radial excitations are important in Landau damping.  If we calculated the the damping rate ignoring the radial thermal excitations, the damping rate would be order of magnitude smaller.
This means that the radial thermal excitations make significant contributions to the Landau damping.   

\section{Conclusion}
In this paper, we studied the Landau damping of dipole oscillations of Bose-condensed gases in a combined potential of highly-elongated harmonic trap and 1D optical lattice, with explicitly including the effect of the radial excitations. While we treated the condensate wavefunction only with the lowest radial mode, we took into account the radial excitations for thermal cloud. \par 
First, we studied equilibrium properties of Bose-condensed gases in a combined harmonic trap and 1D optical lattice potentials. We have presented a detailed calculation of the condensate fraction in a 1D optical lattice at finite temperatures. We find the negative shift of the excitation energies with increasing lattice depth. We obtain the dipole-mode frequency as a function of the lattice depth. The negative shift of the dipole-mode frequency was also observed in the experiment.\par
Second, we calculated Landau damping rate of dipole-modes with varying lattice depth and temperature. Result for the damping rate is consistent with experimented data. Therefore, the experimentally observed damping can be understood as Landau
damping. We also showed that the radial thermal excitations are important in both equilibrium condensate fractions and Landau damping rate.\par
\section*{ACKNOWLEDGMENTS}
This research was support by Academic Frontier Project (2005) of MEXT.


\begin{thebibliography}{17}

\bibitem{RMP78}For a recent review, see for example O. Morsch and M. Oberthaler Rev. Mod. Phys. {\bf 78}, 179 (2006)
  
\bibitem{PRL86}{\it{S. Burger et al., Phys. Rev. Lett.}} {\bf 86}, 4447(2001)
\bibitem{Sc293}{\it{F. S. Cataliotti et al., Science}} {\bf  293}, 843(2001)
\bibitem{PRA66}{\it{F. Ferlaino et al., Phys. Rev. A }} {\bf 66}, 011604(2002)  

\bibitem{EPJD27}{\it{M. Kr\"{a}mer et al.,} {Eur. Phys. J. D}} {\bf 27}, 247 (2003)

\bibitem{JLTP116}{\it{E. Zaremba, T. Nikuni and A. Griffin, J. Low. Temp. Phys.}}{\bf 116} 277 (1999)


\bibitem{PRA57}{\it{S.Giorgini, Phys. Rev. A}} {\bf 57}, 2949(1998)
\bibitem{PRL235}{\it{Pitaevskii,L.P.and Stringari,S.} \it{Phys. Rev. Lett.}}
 {\bf 235}, 398 (1997)
\bibitem{PRA61}{\it{M. Guilleumas and L. P. Pitaevskii, Phys. Rev. A}} {\bf 61} 013602 (2000)
\bibitem{NJP5}{\it{B. Jackson and E. Zaremba} {New J. Phys.}} {\bf 5} 88(2003)

\bibitem{PRA 70}{\it{S. Tsuchiya and A. Griffin,} {Phys. Rev. A}} {\bf{70}}, 023611(2004)
\bibitem{PRA 72}{\it{ A. M. Rey, et al., Phys. Rev. A}} {\bf 72}, 033616(2005)
\bibitem{PRA 73}{\it{B. G. Wild, et al., Phys. Rev. A}} {\bf 73}, 023604(2006)

\bibitem{cd06}{\it{E. Arahata and T. Nikuni  J. Low Temp. Phys. }}{\bf 148}, 345 (2007) 

\bibitem{PRB53}{\it{A. Griffin, Phys. Rev. B}} {\bf 53}, 9341(1995)


\bibitem{PRA69}{\it{P.B.Blakie and J.V.Porto, Phys. Rev. A}} {\bf 69}, 013603(2004)


 \bibitem{p/s}{\it{C. J. Pethich and H. Smith,} {Bose-Einstein Condensation in Dilute Bose Gass}} (UNIVERSITY PRESS CAMBRIDGE)

\end{thebibliography}
\end{document}